\documentclass[prd,aps,a4paper,twocolumn]{revtex4}

\usepackage{graphicx,psfrag}
\usepackage{mathrsfs}

\begin{document}


\title{Summary of GR18 Numerical Relativity parallel sessions (B1/B2
  and B2), Sydney, 8-13 July 2007}

\author{Carsten Gundlach}

\affiliation{School of Mathematics, University of Southampton,
Southampton, SO17 1BJ, UK}

\begin{abstract}
The numerical relativity session at GR18 was dominated by physics
results on binary black hole mergers. Several groups can now simulate
these from a time when the post-Newtonian equations of motion are
still applicable, through several orbits and the merger to the
ringdown phase, obtaining plausible gravitational waves at infinity,
and showing some evidence of convergence with resolution. The results
of different groups roughly agree. This new-won confidence has been
used by these groups to begin mapping out the (finite-dimensional)
initial data space of the problem, with a particular focus on the
effect of black hole spins, and the acceleration by gravitational wave
recoil to hundreds of $km/s$ of the final merged black hole. Other
work was presented on a variety of topics, such as evolutions with
matter, extreme mass ratio inspirals, and technical issues such as
gauge choices.
\end{abstract}

\maketitle


\section{Introduction}

In the numerical simulation of comparable-mass black hole binary
mergers, after a decade of struggle with unstable codes, several
groups are now obtaining reliable gravitational wave signals, and are
competing closely to investigate the parameter space.

This topic dominated the plenary talk by B. Br\"ugmann, as well as
about half of the talks submitted to the parallel session B2 Numerical
Methods. To celebrate the fact that binary black hole simulations are
at last obtaining astrophysics results, a session was held jointly
with the B1 Relativistic Astrophysics parallel session. 7 contributed
talks were given in B1/B2 and 24 more in B2. There were 6 posters in
B2.

Much work is also being done on numerical simulations of astrophysical
scenarios involving matter in general relativity, in particular core
collapse and neutron star binary mergers, but at GR18 this interesting
and active area was represented by few contributed talks. There were
no talks on computer algebra.

Given the remarkable sudden progress in binary black hole evolutions
since 2005, and the remarkable similarity of the results from
different groups, I begin with an overview of this field. I have aimed
to make the astrophysical and historical overviews
(Secs.~\ref{sec:astrophysics} and \ref{sec:history}) accessible to
researchers outside numerical relativity, and a description of the
state of the art (Sec.~\ref{sec:stateoftheart}) useful to beginning
graduate students in the field.  After that, the summaries of
individual talks are likely to be too technical to be accessible to
any but the specialists, and too short to tell the specialists
anything new. Therefore, where I could identify a paper or e-print
related to a specific talk, I give that reference.

\section{Black hole binary mergers}

\subsection{Astrophysical background}

\label{sec:astrophysics}

Many stars are in binary systems, and many of these are expected to
consist, at the end of their life, of two black holes, two neutron
stars, or a neutron star and a black hole. All binary systems lose
orbital energy through emitting gravitational radiation, and merge
eventually, although the time scale on which this happens depends on
their initial separation and masses. Compact objects such as neutron
stars and black holes approach each other very closely before merger,
which means that they can emit significant amounts of gravitational
radiation just before merger, releasing an energy up to a few percent
of the total mass of the system. 

These gravitational waves are very much harder to detect than the same
energy in light, basically because the frequencies are very much
lower. This has the twin consequences that the instrument has to be
very much larger, and that, unlike a CCD camera for light, it cannot
detect individual quanta. For these reasons, no gravitational waves
have yet been measured directly. (However, the time series of several
pulsars in close binaries agrees very accurately with the energy loss
through gravitational waves predicted by general relativity.)

Mergers of compact object binaries are expected to be the principal
source of gravitational waves with frequencies of tens to thousands of
Hertz. The large interferometric detectors LIGO, GEO, VIRGO and TAMA
are now observing in the hundreds of Hertz frequency; they were
reviewed in the plenary talk of S. Whitcomb. LIGO is funded for an
upgrade that should allow it to see binary mergers in the local group;
this should give it enough sources to see something. (See also the
plenary talks by D. Shadock on space-based detectors, and by
M. A. Papa on gravitational wave astronomy.)

In the mid 1990s, the binary black hole problem seemed the natural
problem for numerical relativity to tackle as soon as computers were
big enough to allow simulations in 3D: it is entirely described by
smooth solutions of the Einstein equations, without the need to model
matter. While the two black holes are still far apart each is
approximately a Kerr black hole, and is parameterised by only its mass
and spin, and their orbit is parameterised by only its ellipticity and
size. The gravitational field itself has an infinite number
of degrees of freedom, but after a few orbits the gravitational wave
field is effectively determined by the history of the orbital
motion. This means that the parameter space of binary mergers which
start sufficiently far apart and where accretion into the black holes
can be neglected is effectively finite-dimensional. 

\subsection{History of numerical simulations}

\label{sec:history}

Ten years and perhaps 500 person-years later, the problem had not been
solved because the numerical simulations were stopped by numerical
instabilities that could not be overcome by just using more
resources. In hindsight, there were overlapping problems concerning
the mathematical formulation which interacted with, and
were sometimes confused with, problems of numerical
discretisation.
These were mainly of two types: the use of ill-posed
formulations of the initial-boundary value problem for general
relativity, and gauge choices not appropriate to the symmetries of the
problem. The evolution time before the code broke down was gradually
extended, with the first full orbit achieved by Br\"ugmann in 1997
\cite{Bruegmann}, but only a very few orbits became possible until
2005.

A first stable simulation with an essentially unlimited number of
orbits up to and through the merger was announced by F. Pretorius in
March 2005 \cite{Pretorius}. He solved the Einstein equations in
modified harmonic coordinates. Their principal part is then a set of
wave equations with characteristics on the physical
lightcones. However, some lower-order terms were modified in order to
modify the harmonic time slicing. Other key elements of his success
include, but were not limited to, imposing outer boundary conditions
by compactifying the numerical domain at spatial infinity, the
excision of a spacetime region inside each black hole (singularity
excision), the use of adaptive mesh refinement, and damping the
constraints by friction-like lower-order terms.

In November 2005, an unlimited number of orbits plus merger was
achieved again with very different methods, at the same time but
independently by groups at the University of Texas/Brownsville
\cite{Brownsville} and at NASA/Goddard \cite{Goddard}. These groups
used the ``BSSN'' formulation of the Einstein equations \cite{BSSN},
which had already been successful in neutron star and core collapse
simulations (see for example \cite{Shibata}). The key additional
ingredient for binary black hole simulations was a new version of an
old method for dealing with the singular interior of the black holes
called the ``puncture method'' \cite{BrandtBruegmann}, where inside
each black hole is a wormhole, rather than a collapsed star.

\subsection{State of the art}

\label{sec:stateoftheart}

These two formulations of the problem are still the dominant ones
today with only minor changes. The BSSN formulation is now always used
together with a ``Bona-Mass\'o 1+log'' type slicing condition and a
``hyperbolic $\Gamma$-driver'' type shift condition. With this gauge,
BSSN is hyperbolic, although some of its characteristics are spacelike
and some timelike with respect to the physical light cones. There
seems to be a trend to gauge drivers with time derivatives in the
direction normal to the slice, rather than along the lines of constant
spatial coordinates. For example, $(\partial_t +
\beta^i\partial_i)\alpha = - 2 \alpha K$, rather than $\partial_t
\alpha= - 2 \alpha K$, and similarly for the evolution of the
shift. This avoids a breakdown of hyperbolicity when gauge speeds
coincide, and allows the final black hole to settle down to a Killing
coordinate system.

Encouraged by Pretorius' success, other groups are now experimenting
with modified harmonic coordinates both for vacuum and matter
evolutions, and his work seems to have influenced even the evolution
of the linearised Einstein equations for modelling extreme mass ratio
inspirals, such as stellar mass black holes falling into supermassive
black holes.

The most reliable initial data now seem to be produced by converting a
snapshot of a post-Newtonian (PN) orbit into ``puncture'' type initial
data for full general relativity by solving the Hamiltonian and
momentum constraint. This PN orbit is typically constructed by calculating the
adiabatic shrinking of a conservative PN orbit. To reduce
eccentricity, the post-Newtonian equations of motion can be evolved
over hundreds of orbits until the eccentricity has been radiated
away. It is puzzling that the conversion works so well given that this also
involves a translation of gauge from the post-Newtonian to the full
general relativity framework.

Outer boundaries remain primitive - some in common use are not known
to lead to well-posed initial-boundary value problems, and are known
not to be compatible with the constraint equations. Nevertheless,
numerical boundary conditions exist that are empirically stable, and
the strategy of moving them as far out as possible, typically by using
nested boxes of Cartesian coordinate grids, seems to work. The location
of the outer boundary does not seem to seriously affect the
gravitational wave signals.

Waves are extracted on an approximately spherical surface either by
finding the Zerilli gauge-invariant linear perturbation with respect
to a Schwarzschild background, or by constructing an appropriate
tetrad and calculating the Newman-Penrose scalar $\Psi_4$. Here it
does seem to be important to push this sphere as far out as
possible. 

Black hole masses and spins are calculated by finding a dynamical
horizon, and constructing approximate Killing vectors for use in
Komar-type integrals, by energy and angular momentum balance
arguments, or by fitting quasinormal ringdowns to the known ringdown
of Kerr. Approximate ADM quantities integrated over a large sphere
also seem to agree with these diagnostics.

\subsection{Contributed talks} 

The contributed talks were dominated by physics results which appear
to agree between research groups. Most talks focused on the effect of
the spin of the two black holes on gravitational recoil
(``kicks''). It appears that velocity of several hundred $km/s$ are
generic, and several thousand $km/s$ are possible. This is
astrophysically relevant as it may completely eject the merged black
hole from its galaxy. Note that as velocity is naturally measured in
units of the speed of light, and vacuum gravity is scale-invariant,
these velocities are invariant under an overall scaling of the initial
data. Therefore the results reported here would apply equally to
stellar mass and supermassive black hole binaries, as long as the two
masses are of the same order of magnitude. 

P. Diener \cite{Diener} reported on joint work by the AEI and LSU
groups. Initial data were quasi-circular, determined by an effective
potential method. Kicks up to $175km/s$ were achieved for non-spinning
unequal mass binaries (mass ratio $0.36$), while kicks up to $440km/s$
were obtained with equal mass black holes with spins anti-aligned
parallel to the orbital angular momentum. The post-Newtonian analysis
of gravitational recoil by Kidder \cite{Kidder1995} was found to be
qualitatively valid beyond its domain of applicability.

M. Hannam reviewed the work of the Jena group. They had concentrated
on a configuration with equal masses, and spins anti-aligned in the
orbital plane (``superkicks'') \cite{Hannamkicks}. The magnitude
depended sinusoidally on the angle of the spins in the orbital
plane. Because of the high symmetry, most of the energy was radiated as
$l=2,m=\pm 2$ waves, and the kick could be estimated through the energy
difference between these. The velocities obtained were of the order of
$2500 km/s$ (for $J=0.723M^2$), in agreement with what had been found
by the Brownsville group \cite{Brownsville} (which was not
represented at GR18).

Work was also under way on a bank of numerical template wave forms for
non-spinning mergers \cite{Hannamtemplates}. Full numerical evolutions
based on post-Newtonian initial data could be compared to a
continued post-Newtonian evolution for up to 9 more orbits before the
merger, and agreed very accurately. Going to sixth order accuracy in
the spatial finite differencing had been crucial to obtain a small
enough phase error (2 degrees over 9 orbits) for this. Moreover, very
accurate ``hybrid'' wave forms could be constructed by matching a
post-Newtonian inspiral to black hole ringing, using information from
numerical simulations to match them. This might be used to fill the
parameter space with templates.

P. Laguna \cite{Laguna} reported work from the PSU group on spins
parallel to the orbital angular momentum (up to $400km/s$) and
superkick configurations, and stressed that the Kidder post-Newtonian
formula for spin-orbit interactions, although out of the domain of its
validity, could be used as a heuristic formula with a small number of
free parameters to be determined by full numerical
simulations. Quasi-circular orbits were compared to post-Newtonian
ones over 9 orbits, and agreed, although to lower precision than that
obtained by the Jena group. The moving punctures method was
robust. Waveform comparisons between groups were now needed. 
 
B. Kelly presented the work of the Goddard group. Hybrid waveforms
agreed accurately with fully numerical ones. The code was fifth-order
accurate. The effects of unequal masses and spins were being
analysed. He described an analysis of the gravitational wave recoil in
multipoles \cite{Kellykicks}, with the force seen as a sum of mainly
three products of pairs of multipoles, and stressed that the integral
of momentum over time was not monotonic (``antikicks''). This model
could also explain why spins in the orbital plane produced larger kicks.
However, accretion would tend to align spins with orbital angular
momentum. Finally, the effective one-body approach was a step forward
in covering the parameter space with approximate wave forms.

M. Scheel \cite{Scheel} presented work by the Caltech group that uses
a modified harmonic formulation. (In contrast to Pretorius, their
formulation is reduced to first order in space and time, and
consistent boundary conditions are applied at finite radius. In
contrast to all other groups except the Meudon group they use spectral
methods rather than finite differencing in space.) Currently their
code experienced coordinate problems during merger, but spectral
methods allowed for the currently most accurate inspiral simulations in
full general relativity. Post-Newtonian based initial data (equal
mass, no spin, zero eccentricity) evolved for 30 orbits until merger
agree accurately for the first 15 orbits, and then significant
disagreement could be shown, after taking into account the different
gauges of what is being compared. It was found that quasi-circular
initial data are actually slightly eccentric.

F. Pretorius \cite{Pretoriuscrit} was interested in unusual threshold
behaviour in binary black hole merger, rather than astrophysical wave
forms. For non-spinning black holes of comparable mass, at the
threshold of immediate merger, prompt merger was delayed and replaced
by a whirling phase in which the number of orbits $n$ depends on the
impact parameter $b$ as $\exp n \sim |b-b_*|^\gamma$. Although fully
nonlinear, this behaviour was also shown by point particle geodesics
in Kerr. During the whirl, 1-1.5\% of the energy was radiated per
orbit. He also offered an explanation of superkicks in terms of each
black hole experiencing frame dragging by the other. He expected no
more surprises in generic orbits, except perhaps in the limit of
extreme rotation, or infinite boost.

D. Pollney \cite{Pollney} presented work of the AEI and LSU group on
spin interactions. Distinctive features in the wave forms could be
associated with the spins. The momentum flux could be estimated in
particular by the terms $Q_{22}^+Q^+_{3-3}$ and
$Q^+_{2-2}Q_{21}^\times$.
 
D. Shoemaker \cite{Shoemaker} presented a quantitative study by the
PSU group of the influence of spurious radiation in the initial data
on the merger, by artificially adding $l=m=2$ Teukolsky waves inside
the binary. This changed the ADM mass but not the angular momentum,
with the merger and ringdown relatively unaffected. 

I. Hinder from the PSU group focused on the comparison with
post-Newtonian evolutions over 9 orbits to merger. Quasi-circular
puncture initial data actually showed slight
eccentricity. Disagreement with PN evolutions became more noticeable
in the last 3 orbits. The gravitational waves matched the
post-Newtonian ones well, although the eccentricity showed, as well as
a dependence on resolution towards the end. The wave form still
converged to 3rd to 4th order until the merger.

S. Husa \cite{Husa} from the Jena group spoke on a variety of topics arising in
evolutions. Higher order finite differencing was needed for accuracy
and, surprisingly, seemed to work well with punctures, which are not
smooth. 4th order Kreiss-Oliger dissipation was important, as well as
using an advection stencil for the shift terms. Phase error converged
to 6th order, but not amplitude error; this could be fixed by
re-parameterising by phase. Very low eccentricity in the initial data
(obtained by evolving the post-Newtonian equations over many orbits)
was needed to see any disagreement between full numerical and 3.5PN
evolutions.

J. Gonzalez, also at Jena, reported on unequal mass mergers, with a
mass ratio of up to 10. Gauge problems arose which may be related to
the damping term in the shift driver. The kick speeds agreed well with
\cite{Gonzalez}. As the mass ratio increased, less energy was radiated
in $l=2$, and more in $l=3,4,5$.

W. Tichy \cite{Tichy} reported on work at FAU on binaries with
$J=0.8M^2$, resulting in kicks of up to $2500km/s$. With 10 levels
of mesh refinement, evolutions with an outer boundary at $240M$ could
be run on a 32Gb workstation. The largest error seemed to come from
the extraction radius.

\section{Contributed talks on other topics}

\subsection{Physics results} 

R. de Pietri \cite{dePietri} described simulations of instabilities in
rapidly rotating stars with a $\Gamma$-law equation of state
(initially polytropic, and no shocks form). The threshold of
instability was well approximated by a Newtonian analysis. A bar ($m=2$
dominated) appeared initially, but later resolved into an $m=1$ or
$m=3$ dominated structure.

S. Liebling described the application of a parallel adaptive mesh
refinement code in harmonic coordinates to different physics
problems. Binary boson stars \cite{Lieblingboson} could usefully be
compared with binary black holes to see how the gravitational waves
emitted depend on the internal structure. Here, the relative phase of
the two stars provides an additional arbitrary parameter. Neutron star
binaries had also been simulated \cite{Lieblingneutron}. The old
problem of the critical collapse of Brill waves had been taken up
again, although the critical amplitude had only been approached to
within 1\%. Subcritical scaling gave $\gamma\simeq 0.23$,
$\Delta\simeq 0.75$, which is quite different from the original
results \cite{AbrahamsEvans} $\gamma\simeq 0.36$, $\Delta\simeq
0.60$. Magneto-hydrodynamics had been implemented \cite{LieblingMHD}.

U. Sperhake \cite{Sperhake} reported simulations of head-on collisions
of black holes which had different internal structure, namely
Brill-Lindquist (BL) initial data (each black hole is a wormhole to a
separate internal infinity), Misner data (both black holes connect to
the same internal infinity) and approximate initial data obtained by
superposing two Kerr-Schild (KS) slices through Schwarzschild (where
the slices end at the future spacelike singularity). (His is the only
BSSN code that can stably move excised black holes, which is necessary
for evolving these, non-puncture initial data. Pretorius also uses
moving excision, but in the harmonic formulation). KS had much more
spurious initial radiation. BS and Misner agreed closely, but there was
a small discrepancy between these and KS in the wave form which might be
due to numerical error.

B. Kol reviewed a number of problems in higher dimensions, often
string-inspired, which might usefully be investigated by numerical
relativity. In particular he speculated on a possible relation between
Choptuik's critical solution in the gravitational collapse of a scalar
field in 4D, and the pinching off of a black string in
$R^{d-1,1}\times S^1$ \cite{Kol}.

\subsection{Gauge and boundaries}

The talks by D. Brown \cite{Brown} and D. Garfinkle \cite{Garfinkle}
were closely related to the talk of S. Husa \cite{Husagauge} given in
parallel session A3: Why does the ``moving punctures'' approach to
representing black holes work in numerical relativity? The initial
data \cite{BrandtBruegmann} represent a wormhole opening out to
another universe in the Kruskal solution. However, evolving with the
slicing condition now used by most binary black hole groups (``Bona-Mass\'o
1+log''), numerical error allowed the solution inside the black hole to
jump to a slicing of the Kruskal spacetime where the wormhole is
asymptotically cylindrical and ends at the internal future null
infinity, rather than at the internal spacelike infinity. The slicing
could then become Killing both inside and outside the black hole. At the
same time a ``$\Gamma$-driver'' type shift condition removed grid
points from the interior, producing in effect excision by
under-resolution.

There were two other talks on coordinate choices for black hole
evolutions. D. Hilditch \cite{Hilditch} discussed slicings of
asymptotically flat spacetimes that are asymptotically null combined
with radial compactification, such that the speed of outgoing waves
remains constant. Numerical tests of the spherical wave equation on
Schwarzschild resolved outgoing waves to arbitrarily large radii on a
small number of grid points. L. Lindblom discussed generalisations of
the ``harmonic coordinate driver'' coordinate conditions used by
Pretorius.

J. Seiler presented an implementation by the AEI group of boundary
conditions for the generalised harmonic formulation of the Einstein
equations which are at once compatible with the constraint and result
in a well-posed initial-boundary value problem
\cite{KreissWinicour}. These had been implemented using summation by
parts techniques for finite differencing and the code was now being
tested.

L. Brewin \cite{Brewin} reviewed his work on smooth lattice numerical
relativity: in contrast to Regge calculus, spacetime is not locally
flat, and the legs of the simplices are geodesics.

\subsection{Evolution codes}

J. Thornburg review ongoing work on an adaptive mesh refinement code
in double null coordinates for the linearised vacuum Einstein
equations in harmonic gauge. The code has been tested with the scalar
self-force on circular orbits in Schwarzschild \cite{Barack}. It is to
be applied to the calculation of the self-force on generic particle
orbits in the Schwarzschild, and later Kerr spacetime.

B. Zink reviewed progress on the application of high order,
multipatch, summation-by-parts finite differencing techniques
\cite{multipatch} to the simulation of spacetimes with perfect fluid
and magnetohydrodynamics matter. Initial applications are to a
rotating black hole with an accretion disk. A particular strength of
the multipatch approach was that it allowed the combination of high
resolution in the radial direction with low resolution in the angular
directions.

J. Novak \cite{Novak} reviewed a fully constrained nonlinear 3D evolution code
where only two variables are evolved: in a suitable linearised limit
these represent the two polarisations of gravitational waves. The
spatial gauge is Dirac gauge, defined with respect to a flat
background metric. The code uses spherical coordinates and spectral
methods. Boundary conditions are imposed at finite distance which are
absorbing for quadrupolar waves on Minkowski. It was planned to evolve
a single black hole with horizon boundary conditions, and to include matter. 

\subsection{Initial data}

J. Read \cite{Read} reported on numerical work in the helical Killing
field approximation to the binary inspiral problem, which could be
seen as an alternative to the waveless (conformally flat)
approximation.  A scalar field toy model and a neutron star binary
had been implemented. The equations are mixed
hyperbolic-elliptic. Convergence at present required the outer
boundary to be very close in, but this might be improved by an outer
patch where $\partial_t h_{ij}=0$ is imposed instead.

B. Kelly \cite{Kelly} presented initial data of the puncture type in
full general relativity but based on a snapshot of a post-Newtonian
binary orbit, and with approximately correct gravitational wave
content also based on that orbit. The constraints are currently not
solved, but are violated only at $O[(v/c)^5]$.

\subsection{Analysis}

J. L. Jaramillo \cite{Jaramillo} described the application of a
refined Penrose inequality in axisymmetry, namely $A\le
8\pi(M^2+\sqrt{M^4+J^2})$, as a geometric test for a new apparent
horizon finder \cite{LinNovak}, and for the outermost character of
marginal trapped surfaces.  Numerical evidence of Dain's proposal for
characterizing Kerr data by the saturation of such inequality was also
presented.

G. Cook \cite{Cook} presented methods for finding an approximate
rotational Killing vector in numerical evolutions, which could then be
used in a Komar-type integral to compute the spin of a black
hole. These are constructed by minimising an appropriated measure of
the residue in the Killing equation, resulting in a system of coupled
elliptic equations. This was being tested, and might provide a new tool
for analysing the horizon geometry.

N. Bishop presented progress on the extraction of gravitational waves
at scri by matching to an outgoing null grid. Problems with numerical
noise had been overcome, and the extraction had been tested in the
head-on collision of two black holes in a harmonic code with
constraint-preserving boundary conditions. Full Cauchy-characteristic
matching now seemed feasible. 

A. Norton \cite{Norton} presented animations of a spatial
coordinate system that rotates with a rotating star in a central
region but is fixed at infinity, without getting increasingly
entangled: it is based on the double cover of $SO(3)$ by $S^3$, and so
returns to its original state after two rotations.


\acknowledgments

I would like to thank I. Hinder for comments on the manuscript. 


\end{document}